\documentclass[a4paper]{article}
\usepackage{arxiv}
\pdfoutput=1

\usepackage[T1]{fontenc}
\usepackage[utf8]{inputenc}

\usepackage{booktabs}
\usepackage{makecell}
\usepackage{relsize}
\usepackage{amssymb}
\usepackage[binary-units=true]{siunitx}
\usepackage[labelformat=parens]{subcaption}

\usepackage{siunitx}
\usepackage{relsize}

\newcommand{\astr}[1]{\renewcommand{\arraystretch}{#1}}
\newcommand{\ie}{i.e.}
\newcommand{\eg}{e.g.}
\newcommand{\ea}{et\ al.}
\newcommand{\bigO}{\mathcal{O}}

\newcommand{\Kilo}{\ensuremath{\mathrm{k}}}
\newcommand{\Mega}{\ensuremath{\mathrm{m}}}
\newcommand{\AlgLongName}[1]{\textit{#1}}

\newcommand{\VertexSet}{V}

\newcommand{\initE}{m}
\newcommand{\avgE}{\overline{m}}
\newcommand{\avgD}{\overline{d}}
\newcommand{\finE}{M}
\newcommand{\tcE}{m^*}
\newcommand{\maxOutDegree}{\Delta^+}
\newcommand{\Arc}[2]{(#1, #2)}
\newcommand{\Degree}[1]{\ensuremath{\mathrm{deg}(#1)}}
\newcommand{\InDegree}[1]{\ensuremath{\mathrm{deg}^\mathsmaller{-}(#1)}}
\newcommand{\OutDegree}[1]{\ensuremath{\mathrm{deg}^\mathsmaller{+}(#1)}}

\newcommand{\Ratio}{\rho}

\newcommand{\List}[1]{\ensuremath{\mathcal{L}}}
\newcommand{\InNeighborList}[1]{\ensuremath{\mathcal{N}^-}}
\newcommand{\OutNeighborList}[1]{\ensuremath{\mathcal{N}^+}}
\newcommand{\InEdgesList}[1]{\ensuremath{\mathcal{E}^-}}
\newcommand{\OutEdgesList}[1]{\ensuremath{\mathcal{E}^+}}

\newcommand{\Queue}[1]{\Code{Q}}

\newcommand{\Ops}{\sigma}

\newcommand{\Updates}{\delta}
\newcommand{\AddsRatio}{\Updates{}_{\mathsmaller{+}}}

\newcommand{\SCC}{\text{SCC}}
\newcommand{\MatrixMult}{\omega}
\newcommand{\DAG}{DAG}

\newcommand{\OutReachSet}{{R^\mathsmaller{+}}}
\newcommand{\InReachSet}{{R^\mathsmaller{-}}}

\newcommand{\SSoSiR}{SSR}

\newcommand{\SDFS}{\texttt{DFS}}
\newcommand{\SBFS}{\texttt{BFS}}
\newcommand{\SIM}{\texttt{SI}}

\newcommand{\SES}{\texttt{SES}}

\newcommand{\RestartLimit}{\ensuremath{\beta}}
\newcommand{\RestartRatio}{\ensuremath{\Ratio}}

\newcommand{\SV}{\texttt{SV}}
\newcommand{\SVA}{\texttt{SVA}}
\newcommand{\SVC}{\texttt{SVC}}
\newcommand{\SVp}[1]{\texttt{SV(#1)}}
\newcommand{\SVAp}[1]{\texttt{SVA(#1)}}
\newcommand{\SVCp}[1]{\texttt{SVC(#1)}}
\newcommand{\FBSBFS}{\texttt{BiBFS}}
\newcommand{\SDBFS}{\texttt{DBFS}}

\newcommand{\Code}[1]{\texttt{#1}}

\newcommand{\Routine}[1]{\textsf{#1}}

\newcommand{\Tool}[1]{\textsf{#1}}
\newcommand{\KONECT}{\Tool{KONECT}}
\newcommand{\SNAP}{\Tool{SNAP}}
\newcommand{\ALGORA}{\Tool{Algora}}

\newcommand{\TabLabel}[1]{\label{tab:#1}}
\newcommand{\Table}[1]{Table~\ref{tab:#1}}

\newcommand{\Figure}[1]{Figure~\ref{fig:#1}}
\newcommand{\Figures}[2]{Figures~\ref{fig:#1} and~\ref{fig:#2}}
\newcommand{\FiguresSub}[2]{Figures~\ref{fig:#1},\,\subref{fig:#2}}

\newcommand{\SectLabel}[1]{\label{sect:#1}}
\newcommand{\Section}[1]{Section~\ref{sect:#1}}

\usepackage{color}

\usepackage{tikz}
\usetikzlibrary{matrix}
\usepackage{pgfplots}

\newcommand{\LabelEnumi}[1]{%
\textcolor{cyan}{#1}%
}

\title{Faster Fully Dynamic Transitive Closure in Practice}

\author{Kathrin Hanauer}{University of Vienna, Faculty of Computer Science, Vienna, Austria}{kathrin.hanauer@univie.ac.at}{https://orcid.org/0000-0002-5945-837X}{}
\author{Monika Henzinger}{University of Vienna, Faculty of Computer Science, Vienna, Austria}{monika.henzinger@univie.ac.at}{https://orcid.org/0000-0002-5008-6530}{}
\author{Christian Schulz}{University of Vienna, Faculty of Computer Science, Vienna, Austria}{christian.schulz@univie.ac.at}{https://orcid.org/0000-0002-2823-3506}{}
\authorrunning{K.\ Hanauer, M.\ Henzinger, C.\ Schulz}
\Copyright{Kathrin Hanauer, Monika Henzinger, Christian Schulz}

\keywords{Dynamic Graph Algorithms, Reachability}

\funding{The research leading to these results has received funding
from the European Research Council under the European Union’s
Seventh Framework Programme (FP/2007-2013) / ERC Grant
Agreement no.\ 340506.}%

\date{}

\begin{document}
\maketitle
\begin{abstract}
The fully dynamic transitive closure problem asks to maintain reachability
information in a directed graph between arbitrary pairs of vertices, while the
graph undergoes a sequence of edge insertions and deletions. The problem has
been thoroughly investigated in theory and many specialized algorithms for solving
it have been proposed in the last decades. In two large studies [Frigioni
ea, 2001; Krommidas and Zaroliagis, 2008], a number of these algorithms have
been evaluated experimentally against simple static algorithms for graph
traversal, showing the competitiveness and even superiority of the simple
algorithms in practice, except for very dense random graphs or very high ratios
of queries. A major drawback of those studies is that only small and mostly
randomly generated~graphs~are~considered.

In this paper, we engineer new algorithms to maintain all-pairs reachability
information which are simple and space-efficient.
Moreover, we perform an extensive experimental evaluation on
both generated and real-world instances that are several orders of magnitude
larger than those in the previous studies. Our results indicate that our new algorithms
outperform all state-of-the-art algorithms on all types of input considerably in practice.
\end{abstract}

\section{Introduction}
Complex graphs are useful in a wide range of applications from technological networks to biological systems like the human brain.
These graphs can contain billions of vertices and edges. Analyzing these networks aids us in gaining new insights about our surroundings.
One of the most basic questions that arises in this setting is whether one vertex can \emph{reach} another vertex via a directed path.
This simple problem has a wide range of applications such program
analysis~\cite{reps1998program}, protein-protein interaction networks~\cite{10.1093/nar/gkq1207}, centrality measures, and is used as subproblem in a wide range of more
complex (dynamic) algorithms such as in the computation of (dynamic) maximum flows~\cite{fordfulkerson1956,Edmonds1972,goldberg2011maximum}.
Often, the underlying graphs or input instances change over time, \ie, vertices
or edges are inserted or deleted as time is passing.
In a social network, for example, users sign up or leave, and relations between
them may be created or removed over time.
Terminology-wise, a problem is said to be \emph{fully dynamic} if the update
operations include both insertions \emph{and} deletions of edges, and
\emph{partially dynamic} if only one type of update operations is allowed.
In this context, a problem is called \emph{incremental}, if only edge
insertions occur, but no deletions, and \emph{decremental} vice versa.

Recently, we studied an extensive set of algorithms for the
\emph{single-source} reachability problem in the fully dynamic
setting~\cite{HHS20}.
The fully dynamic single-source reachability problem maintains the set of
vertices that are reachable from a given \emph{source vertex}, subject to edge
deletions and insertions.
In particular, we designed several fully dynamic variants of well-known
approaches to obtain and maintain reachability information with respect
to~a~distinguished~source.

This yields the \emph{starting point of this paper}: our goal was to
transfer recently engineered algorithms for the fully dynamic
\emph{single-source} reachability
problem~\cite{HHS20} to the more general fully
dynamic \emph{transitive closure} problem (also known as fully dynamic
all-pairs reachability problem).
In contrast to the single-source problem, the \emph{fully dynamic transitive
closure} problem consists in maintaining reachability information between
\emph{arbitrary} pairs of vertices $s$ and $t$ in a directed graph, which in
turn is subject to edge insertions and deletions.
If the graph does not change, \ie, in the \emph{static} setting, the question
whether an arbitrary vertex $s$ can reach another arbitrary vertex $t$ can
either be answered in linear time by starting a breadth-first or depth-first
search from $s$, or it can be answered in constant time after the transitive
closure of the graph, \ie, reachability information for all pairs of vertices,
has been computed.
The latter can be obtained in $\bigO(n^\MatrixMult{})$, where $\MatrixMult{}$
is the exponent in the running time of the best-known fast matrix
multiplication algorithm (currently, $\MatrixMult{} <
2.38$~\cite{Coppersmith90}), or combinatorially in $\bigO(n\cdot m)$ or
$\bigO(n^3)$ time by either starting a breadth-first or depth-first search from
each vertex or using the Floyd-Warshall
algorithm~\cite{Floyd62,Warshall62,CLRS09}.

In the dynamic setting, the aim is to avoid such costly recomputations from
scratch after the graph has changed, especially if the update was small.
Hence, the dynamic version of the problem has been thoroughly studied in theory
and many specialized algorithms for solving it have been proposed in the last
decades.
However, even the insertion or deletion of a single edge may affect the
reachability between $\Omega(n^2)$ vertex pairs, which is why one cannot hope for an
algorithm with constant query time that processes updates in less than
$\bigO(n^2)$ worst-case time if the transitive closure is maintained explicitly.
Furthermore, conditional lower bounds~\cite{HKNS15} suggest that no faster
solution than the na\"{\i}ve recomputation from scratch is possible after each
change in the graph.

Whereas the static approach to compute the transitive closure beforehand via
graph traversal can be readily adapted to the incremental setting, yielding an
amortized update time of $\bigO(n)$~\cite{Italiano86}, a large number of
randomized and deterministic
algorithms~\cite{IK83,Italiano88,HK95,KS02,King99,KT01,DI05,DI08,RZ08,Roditty08,Sankowski04,Lacki11}
has been devised over the last years for the decremental and the fully
dynamic version of the problem.
The currently fastest algorithm in the deletions-only case is deterministic,
has a total update time of $\bigO(n\cdot m)$, and answers queries in constant
time~\cite{Lacki11}.
In the fully dynamic setting, updates can be processed deterministically in
$\bigO(n^2)$ amortized time with constant query time~\cite{DI08}, or,
alternatively in $\bigO(m\sqrt{n})$ amortized update time with
$\bigO(\sqrt{n})$ query time~\cite{RZ08}.
An almost exhaustive set of these algorithms has been evaluated experimentally
against simple static algorithms for graph traversal such as breadth-first or
depth-first search in two large studies~\cite{FMNZ01,KZ08}.
Surprisingly, both have shown that the simple algorithms are competitive and
even superior to the specialized algorithms in practice, except for dense
random graphs or, naturally, very high ratios of queries.
Only two genuinely dynamic algorithms could challenge the simple ones:
an algorithm developed by Frigioni~\ea{}~\cite{FMNZ01}, which is based on
Italiano's algorithms~\cite{Italiano86,Italiano88} as well as an extension of a
decremental Las Vegas algorithm proposed by Roditty and Zwick~\cite{RZ08},
developed by Krommidas and Zaroliagis~\cite{KZ08}.
Both rely on the computation and maintenance of strongly connected components,
which evidently gives them an advantage on dense graphs.
Nevertheless, they appeared to be unable to achieve a speedup of a factor greater
than ten in comparison to breadth-first and depth-first search.

In this paper, we engineer a family of algorithms that build on recent
experimental results for the single-source reachability problem~\cite{HHS20}.
Our algorithms are very easy to implement and benefit from strongly connected
components likewise, although they do not (necessarily) compute them
explicitly.
In an extensive experimental evaluation on various types of input instances,
we compare our algorithms to all simple-minded algorithms from~\cite{FMNZ01,KZ08}
as well as a modified, bidirectional breadth-first search.
The latter already achieves a speedup of multiple factors over the standard
version, and our new algorithms outperform the simple-minded algorithms on all
types of input by several orders of magnitude in practice.
\section{Preliminaries}%
\label{preliminaries}
\textbf{Basic Concepts.}
Let $G = (V, E)$ be a directed graph with vertex set $V$ and edge set
$E$.
Throughout this paper, let $n=|V|$ and $m=|E|$.
The \emph{density} of $G$ is $d = \frac{m}{n}$.
An edge $\Arc{u}{v} \in E$ has \emph{tail} $u$ and \emph{head} $v$
and $u$ and $v$ are said to be \emph{adjacent}.
$\Arc{u}{v}$ is said to be an \emph{outgoing} edge or \emph{out-edge} of $u$
and an \emph{incoming} edge or \emph{in-edge} of $v$.
The \emph{outdegree} $\OutDegree{v}$/\emph{indegree}
$\InDegree{v}$/\emph{degree} $\Degree{v}$ of a vertex $v$ is its number of
(out-/in-) edges.
The \emph{out-neighborhood} (\emph{in-neighborhood}) of a vertex
$u$ is the set of all vertices $v$ such that $\Arc{u}{v} \in E$ ($\Arc{v}{u}\in E$).
A sequence of vertices $s \to \cdots \to t$ such that each pair of consecutive
vertices is connected by an edge is called an \emph{$s$-$t$ path} and $s$ can
\emph{reach} $t$.
A \emph{strongly connected component} (\emph{\SCC{}}) is a maximal subset of
vertices $X \subseteq V$ such that for each ordered pair of vertices $s,t \in
X$, $s$ can reach $t$.
The \emph{condensation} of a directed graph $G$ is a directed graph $G_C$
that is obtained by shrinking every \SCC{} of $G$ to a single vertex, thereby
preserving edges between different \SCC{}s.
A graph is \emph{strongly connected} if it has only one \SCC{}.
In case that each \SCC{} is a singleton, \ie, $G$ has $n$ \SCC{}s, $G$ is said
to be \emph{acyclic} and also called a \emph{\DAG{}} (directed acyclic graph).
The \emph{reverse} of a graph $G$ is a graph with the same vertex set as $G$,
but contains for each edge $\Arc{u}{v} \in E$ the reverse edge $\Arc{v}{u}$.

A \emph{dynamic graph} is a directed graph $G$ along with an ordered sequence
of updates, which consist of edge insertions and deletions.
In this paper, we consider the \emph{fully dynamic transitive closure problem}: %
Given a directed graph, answer reachability queries between arbitrary pairs of
vertices $s$ and $t$, subject to edge insertions and~deletions.

\subparagraph*{Related Work.}
Due to space limitations, we only give a brief overview over related work by
listing the currently best known results for fully dynamic algorithms on
general graphs in \Table{algorithms-related}.
For details and insertions-only as well as deletions-only algorithms, see
Appendix \ref{sect:app:related}.
\begin{table*}[htb]
\caption{Currently best results for fully dynamic transitive closure.
All running times are asymptotic ($\bigO$-notation).}
\TabLabel{algorithms-related}
\centering
\astr{1.2}
\begin{tabular}{cc@{\hspace{2em}}l}
\toprule
Query Time
&
Update Time
&
\\
\midrule
$m$
&
$1$
&
na\"{\i}ve
\\
$1$
&
$n^2$
&
Demetrescu and Italiano~\cite{DI08},
Roditty~\cite{Roditty08}, %
Sankowski~\cite{Sankowski04} %
\\
$\sqrt{n}$
&
$m\sqrt{n}$
&
Roditty and Zwick~\cite{RZ08}
\\
$m^{0.43}$
&
$m^{0.58}n$
&
Roditty and Zwick~\cite{RZ08}
\\
$n^{0.58}$
&
$n^{1.58}$,
&
Sankowski~\cite{Sankowski04}
\\
$n^{1.495}$
&
$n^{1.495}$
&
Sankowski~\cite{Sankowski04}
\\
$n$
&
$m+n\log n$
&
Roditty and Zwick~\cite{RZ16}
\\
$n^{1.407}$
&
$n^{1.407}$
&
van den Brand~\ea{}~\cite{BNS19}
\\
\bottomrule
\end{tabular}
\end{table*}

In large studies, Frigioni~\ea{}~\cite{FMNZ01} as well as Krommidas and
Zaroliagis~\cite{KZ08} have implemented an extensive set of known
algorithms for dynamic transitive closure and compared them to each other as
well as to static, ``simple-minded'' algorithms such as breadth-first and
depth-first search.
The set comprises the original
algorithms in~\cite{Italiano86,Italiano88,Yellin93,HK95,King99,KT01,Roditty08,RZ08,RZ16,DI05,DI08}
as well as several modifications and improvements thereof.
The experimental evaluations on random Erd\H{o}s-Renyí graphs, instances
constructed to be difficult on purpose, as well as two instances based on
real-world graphs, showed that on all instances except for dense random graphs
or a query ratio of more than 65\%
dynamic ones distinctly and up to several factors.
Their strongest competitors were the fully dynamic extension~\cite{FMNZ01} of
the algorithms by Italiano~\cite{Italiano86,Italiano88}, as well as the fully
dynamic extension~\cite{KZ08} of the decremental algorithm by Roditty and
Zwick~\cite{RZ08}.
These two algorithms also were the only ones that were faster than static graph
traversal on dense random graphs, by a factor of at most ten.
\section{Algorithms}\SectLabel{algorithms}
We propose a new and very simple approach to maintain the
transitive closure in a fully dynamic setting.
Inspired by a recent study on single-source reachability, it is based solely on
single-source and single-sink reachability (\SSoSiR{}) information.
Unlike most algorithms for dynamic transitive closure, it does not explicitly
need to compute or maintain strongly connected components---which can be
time-consuming---but, nevertheless, profits indirectly if the graph is strongly
connected.
Different variants and parameterizations of this approach lead to a family of
new algorithms, all of which are easy to implement and---depending on the
choice of parameters---extremely space-efficient.

In \Section{experiments}, we evaluate this approach experimentally against a
set of algorithms that have been shown to be among the fastest algorithms in
practice so far.
Curiously enough, this set comprises the classic simple, static algorithms for
graph traversal, breadth-first search and depth-first search.
For the sake of completeness, we will start by describing the practical
state-of-the-art algorithms, and then continue with our new approach.
Each algorithm for fully dynamic transitive closure can be described by means
of four subroutines:
\Routine{initialize()}, %
\Routine{insertEdge($\Arc{u}{v}$)}, %
\Routine{deleteEdge($\Arc{u}{v}$)}, %
and \Routine{query($s, t$)}, %
which define the behavior during the initialization phase, in case
that an edge~$\Arc{u}{v}$ is added or removed, and
how it answers a query of whether a vertex $s$ can reach a vertex $t$,
respectively.

\Table{algorithms-overview} provides an overview of all algorithms in this section along with
their abbreviations, whereas \Table{algorithms} subsumes their worst-case time and
space complexities.
All algorithms considered are combinatorial and either deterministic or Las
Vegas-style randomized, \ie, their running time, but not their correctness,
may depend on random variables.
\subsection{Static Algorithms}
In the static setting or in case that a graph is given without further
(reachability) information besides its edges, breadth-first search (\emph{BFS})
and depth-first search (\emph{DFS}) are the two standard algorithms to
determine whether there is a path between a pair of vertices or not.
Despite their simple-mindedness and the fact that they typically have no
persistent memory (such as a cache, \eg), experimental
studies~\cite{FMNZ01,KZ08} have shown them to be at least competitive with
both partially and fully dynamic algorithms and even superior on various
instances.

\textbf{BFS, DFS:} We consider both BFS and DFS in their pure versions:
For each \Routine{query($s, t$)}, a new BFS or DFS, respectively, is initiated
from $s$ until either $t$ is encountered or the graph is exhausted.
The algorithms store or maintain no reachability information whatsoever and
do not perform any work in
\Routine{initialize()}, %
\Routine{insertEdge($\Arc{u}{v}$)}, or %
\Routine{deleteEdge($\Arc{u}{v}$)}. %
We refer to these algorithms simply as \SBFS{} and \SDFS{},
respectively.

In addition, we consider a hybridization of \SBFS{} and \SDFS{}, called
\SDBFS{}, which was introduced originally by Frigioni~\ea{}~\cite{FMNZ01} and
is also part of the later study~\cite{KZ08}.
In case of a \Routine{query($s, t$)}, the algorithm visits vertices in DFS
order, starting from $s$, but additionally checks for each vertex that is
encountered whether $t$ is in its out-neighborhood.

\textbf{Bidirectional BFS:}
To speed up static reachability queries even further, we adapted a
well-established approach for the more general problem of finding shortest
paths and perform two breadth-first searches alternatingly:
Upon a \Routine{query($s, t$)}, the algorithm initiates a customary BFS
starting from $s$, but pauses already after few steps, even if $t$ has not
been encountered yet.
The algorithm then initiates a BFS on the reverse graph,
starting from $t$, and also pauses after few steps, even if $s$ has not been
encountered yet.
Afterwards, the first and the second BFS are resumed by turns, always for a few
steps only, until either one of them encounters a vertex $v$ that has already
encountered by the other, or the graph is exhausted.
In the former case, there is a path from $s$ via $v$ to $t$, hence the
algorithm answers the query positively, and otherwise negatively.
We refer to this algorithm as \FBSBFS{} (\AlgLongName{Bidirectional BFS})
and use the respective out-degree of the current vertex in each BFS as step
size, \ie, each BFS processes one vertex, examines all its out-neighbors,
and then pauses execution.
Note that the previous experimental studies \cite{FMNZ01,KZ08} do not consider this algorithm.

\subsection{A New Approach}
\subparagraph*{General Overview.}
Let $v$ be an arbitrary vertex of the graph and let $\OutReachSet(v)$ and
$\InReachSet(v)$ be the sets of vertices reachable from $v$ and that can reach
$v$, respectively.
To answer reachability queries between two vertices $s$ and $t$, we use the
following simple observations (see also \Figure{observations}):

\begin{enumerate}[(O1)]
\item\label{item:in-s-out-t}
If $s\in\InReachSet(v)$ and $t\in\OutReachSet(v)$, then $s$ can reach $t$.
\item\label{item:out-s-nout-t}
If $v$ can reach $s$, but not $t$, \ie, $s\in\OutReachSet(v)$ and
$t\not\in\OutReachSet(v)$, then $s$ cannot reach $t$.
\item\label{item:nin-s-in-t}
If $t$ can reach $v$, but $s$ cannot, \ie, $s\not\in\InReachSet(v)$ and
$t\in\InReachSet(v)$, then $s$ cannot reach $t$.
\end{enumerate}

Whereas the first observation is widely used in several algorithms,
we are not aware of any algorithms making direct use of the others.
Our new family of algorithms keeps a list $\List{}_{SV}$ of length $k$ of
so-called \emph{supportive vertices}, which work similarly to cluster centers
in decremental shortest paths algorithms~\cite{RZ12}.
For each vertex $v$ in $\List{}_{SV}$, there are two fully dynamic data
structures maintaining the sets $\OutReachSet(v)$ and $\InReachSet(v)$,
respectively.
In other words, these data structures maintain single-source as well as
single-sink reachability for a vertex $v$.
We give details on those data structures at the end of this section.
All updates to the graph, \ie, all notifications of
\Routine{insertEdge($\cdot$)} and %
\Routine{deleteEdge($\cdot$)}, %
are simply
passed on to these data structures.
In case of a \Routine{query($s, t$)}, the algorithms first check whether one of
$s$ or $t$ is a supportive vertex itself.
In this case, the query can be answered decisively using the corresponding data
structure.
Otherwise, the vertices in $\List{}_{SV}$ are considered one by one and the
algorithms try to apply one of the above observations.
Finally, if this also fails, a static algorithm serves as fallback to answer
the reachability query.

Whereas this behavior is common to all algorithms of the family, they differ in
their choice of supportive vertices and the subalgorithms used to maintain
\SSoSiR{} information as well as the static fallback algorithm.

Note that it suffices if an algorithm has exactly one vertex $v_i$ from each
\SCC{} $C_i$ in $\List{}_{SV}$ to answer every reachability query in the same
time as a query to a \SSoSiR{} data structure, \eg,
$\bigO(1)$:
If $s$ and $t$ belong to the same \SCC{} $C_i$, then the supportive vertex
$v_i$ is reachable from $s$ and can reach $t$,
so the algorithm answers the query positively in accordance with
observation~\LabelEnumi{(O\ref{item:in-s-out-t})}.
Otherwise, $s$ belongs to an \SCC{} $C_i$ and $t$ belongs to an \SCC{} $C_j$,
$j \neq i$.
If $C_i$ can reach $C_j$ in the condensation of the graph, then also $v_i$ can
reach $t$ and $v_i$ is reachable from $s$,
so the algorithm again answers the query positively in accordance with
observation~\LabelEnumi{(O\ref{item:in-s-out-t})}.
If $C_i$ cannot reach $C_j$ in the condensation of the graph, then $v_i$
can reach $s$, but not $t$
so the algorithm answers the query negatively in accordance with
observation~\LabelEnumi{(O\ref{item:out-s-nout-t})}.
The supportive vertex representing the \SCC{} that contains $s$ or $t$,
respectively, may be found in constant time using a map; however, updating it
requires in turn to maintain the \SCC{}s dynamically, which incurs additional
costs during edge insertions and deletions.

\textbf{Choosing Supportive Vertices.}
The simplest way to choose supportive vertices consists in picking them
uniformly at random from the set of all vertices in the initial graph and never
revisiting this decision.
We refer to this algorithm as \SVp{$k$} (\AlgLongName{$k$-Supportive
Vertices}).
During \Routine{initialize()}, \SVp{$k$} creates $\List{}_{SV}$ by drawing $k$
non-isolated vertices uniformly at random from $\VertexSet$.
For each $v \in \List{}_{SV}$, it initializes both a dynamic single-source as
well as a dynamic single-sink reachability data structure, each rooted at $v$.
If less than $k$ vertices have been picked during initialization because of
isolated vertices, $\List{}_{SV}$ is extended as soon as possible.

Naturally, the initial choice of supportive vertices may be unlucky, which is
why we also consider a variation of the above algorithm that periodically
clears the previously chosen list of supportive vertices after $c$ update
operations and re-runs the selection process.
We refer to this algorithm as \SVAp{$k, c$} (\AlgLongName{$k$-Supportive
Vertices with $c$-periodic Adjustments}).

As shown above, the perfect choice of a set of supportive vertices consists of
exactly one per \SCC{}.
This is implemented by the third variant of our algorithm, \SVC{}
(\AlgLongName{Supportive Vertices with \SCC{} Cover}).
However, maintaining \SCC{}s dynamically is itself a non-trivial task and has
recently been subject to extensive research~\cite{Lacki11,Roditty13}.
Here, we resolve this problem by waiving exactness, or, more precisely, the
reliability of the cover.
Similar to above, the algorithm takes two parameters $z$ and $c$.
During \Routine{initialize()}, it computes the \SCC{}s of the input graph and
arbitrarily chooses a supportive vertex in each \SCC{} as representative
if the \SCC{}'s size is at least $z$.
In case that all \SCC{}s are smaller than $z$, an arbitrary vertex that is
neither a source nor a sink, if existent, is made supportive.
The algorithm additionally maps each vertex to the representative of its
\SCC{}, where possible.
After $c$ update operations, this process is re-run and the list of supportive
vertices as well as the vertex-to-representative map is updated suitably.
However, we do not de-select supportive vertices picked in a previous round if
they represent an \SCC{} of size less than $z$, which would mean to also
destroy their associated \SSoSiR{} data structures.
Recall that computing the \SCC{}s of a graph can be accomplished in
$\bigO(n+m)$ time~\cite{Tarjan72}.
For a \Routine{query($s,t$)}, the algorithm looks up the \SCC{} representative
of $s$ in its map and checks whether this information, if present, is still
up-to-date by querying their associated data structures.
In case of success, the algorithm answers the query as described in the ideal
scenario by asking whether the representative of $s$ can reach $t$.
Otherwise, the algorithm analogously tries to use the \SCC{} representative of
$t$.
Outdated \SCC{} representative information for $s$ or $t$ is deleted to avoid
further unsuccessful checks.
In case that neither $s$ nor $t$ have valid \SCC{} representatives, the
algorithm falls back to the operation mode of \SV{}.
\begin{table*}[tb]
\caption{Algorithms and abbreviations overview.}
\TabLabel{algorithms-overview}
\centering
\astr{1.2}
\begin{tabular}{@{}ll@{\hskip 33pt}ll@{}}
\toprule
Algorithm & Long name & Algorithm & Long name \\
\midrule
\SDFS{} / \SBFS{}
& static DFS / BFS
& \SV{}
& Supportive Vertices
\\
\SDBFS{}
& static DFS-BFS hybrid
& \SVA{}
& Supportive Vertices with Adjustments
\\
\FBSBFS{}
& static bidirectional BFS %
& \SVC{}
& Supportive Vertices with SCC Cover
\\
\bottomrule
\end{tabular}
\end{table*}

\textbf{Algorithms for Maintaining Single-Source/Single-Sink Reachability.}
To access fully dynamic \SSoSiR{} information,
we consider the two single-source reachability algorithms that have been shown
to perform best in an extensive experimental evaluation on various types of
input instances~\cite{HHS20}.
In the following, we only provide a short description and refer the interested
reader to the original paper~\cite{HHS20} for
details.

The first algorithm, \SIM{}, is a fully dynamic extension of a simple
incremental algorithm and maintains a not necessarily height-minimal
reachability tree.
It starts initially with a BFS tree, which is also extended using BFS in
\Routine{insertEdge($\Arc{u}{v}$)} only if $v$ and vertices reachable
from $v$ were unreachable before.
In case of \Routine{deleteEdge($\Arc{u}{v}$)}, the algorithm tries to
reconstruct the reachability tree, if necessary, by using a combination of
backward and forward BFS.
If the reconstruction is expected to be costly because more than a configurable
ratio $\RestartRatio$ of vertices may be affected, the algorithm instead
recomputes the reachability tree entirely from scratch using BFS.
The algorithm has a worst-case insertion time of $\bigO(n+m)$, and, unless
$\RestartRatio = 0$, a worst-case deletion time of $\bigO(n\cdot m)$.

The second algorithm, \SES{}, is a simplified, fully dynamic extension of
Even-Shiloach trees~\cite{ES81} and maintains a (height-minimal) BFS tree
throughout all updates.
Initially, it computes a BFS tree for the input graph.
In \Routine{insertEdge($\Arc{u}{v}$)}, the tree is updated where
necessary using a BFS starting from $v$.
To implement \Routine{deleteEdge($\Arc{u}{v}$)}, the algorithm employs a
simplified procedure in comparison to Even-Shiloach trees, where the BFS level
of affected vertices increases gradually until the tree has been fully adjusted
to the new graph.
Again, the algorithm may abort the reconstruction and recompute the BFS tree
entirely from scratch if the update cost exceeds configurable thresholds
$\RestartRatio$
and $\RestartLimit$.
For constant $\RestartLimit$, the worst-case time per update operation
(edge insertion or deletion) is in $\bigO(n+m)$.

Both algorithms have $\bigO(n+m)$ initialization time, support reachability
queries in $\bigO(1)$ time, and require $\bigO(n)$ space.
We use the same algorithms to maintain single-sink reachability information by
running the single-source reachability algorithms on the reverse graph.
\Table{algorithms} shows the resulting time and space complexities
of our new fully dynamic algorithms for dynamic closure in combination with
these subalgorithms, where $\RestartRatio > 0$ and $\RestartLimit \in \bigO(1)$.

\section{Experiments}%
\SectLabel{experiments}
\subparagraph*{Setup.}
For the experimental evaluation of our approach and the effects of its
parameters, we implemented\footnote{We plan to release the source code publicly.}
it together with all four static approaches
mentioned in \Section{algorithms} in C++17 and compiled the code with GCC 7.4
using full optimization (\texttt{-O3 -march=native -mtune=native}).
We would have liked to include the two best non-static algorithms from the
earlier study~\cite{KZ08}; unfortunately, the released source code is based
on a proprietary algorithm library.
Nevertheless, we are able to compare our new algorithms indirectly to both by
relating their performance to DFS and BFS, as happened in the earlier study.
All experiments were run sequentially under Ubuntu 18.04 LTS with Linux kernel
4.15 on an Intel Xeon E5-2643 v4 processor clocked at \SI{3.4}{GHz}, where each
experiment had exclusive access to one core and could use solely local memory,
\ie, in particular no swapping was allowed.

For each algorithm (variant) and instance, we separately measured the time
spent \emph{by the algorithm} on updates as well as on queries.
We specifically point out that the measured times exclude the time spent on
performing an edge insertion or deletion \emph{by the underlying graph data
structure}.
This is especially of importance if an algorithm's update time itself is
very small, but the graph data structure has to perform non-trivial work.
We use the dynamic graph data structure from the open-source library
\ALGORA{}\cite{Algora}, which is able to perform edge insertions and deletions
in constant time.
Our implementation is such that the algorithms are unable to look ahead in time
and have to process each operation individually.
To keep the numbers easily readable, we use $\Kilo{}$ and $\Mega{}$ as
abbreviations for $\times 10^3$ and $\times 10^6$, respectively.
\subparagraph*{Instances.}
We evaluate the algorithms on a diverse set of random and real-world instances,
which have also been used in~\cite{HHS20} and are
publicly available\footnote{\url{https://dyreach.taa.univie.ac.at}}.

\emph{ER Instances.}
The random dynamic instances generated according to the Erd\H{o}s-Renyí model $G(n, m)$
consist of an initial graph with $n = \num{100}\Kilo{}$ or $n =
\num{10}\Mega{}$ vertices and $m = d \cdot n$ edges, where $d \in [ \num{1.25}
\dots \num{50}]$.
In addition, they contain a random sequence of \num{100}\Kilo{} operations
$\Ops$ consisting of edge insertions, edge deletions, as well as reachability
queries:
For an insertion or a query, an ordered pair of vertices was chosen uniformly
at random from the set of all vertices.
Likewise, an edge was chosen uniformly at random from the set of all edges
for a deletion.
The resulting instances may contain parallel edges as well as loops and each
operation is contained in a batch of ten likewise operations.

\emph{Kronecker Instances.}
Our evaluation uses two sets of size \num{20} each:
\texttt{kronecker-csize} contains instances with $n \approx
\num{130}\Kilo{}$, whereas those in \texttt{kronecker-growing}
have $n \approx \num{30}$ initially and grow to $n \approx \num{130}\Kilo{}$
in the course of updates.
As no generator for dynamic stochastic Kronecker graph exists, the instances
were obtained by computing the differences in edges in a series of so-called
\emph{snapshot graphs}, where the edge insertions and deletions between two
subsequent snapshot graphs were shuffled randomly.
The snapshot graphs where generated by the \texttt{krongen} tool that is part
of the \SNAP{} software library \cite{leskovec2016snap}, using the estimated
initiator matrices given in~\cite{Leskovec2010} that correspond to real-world
networks.
The instances in \texttt{kronecker-csize} originate from ten snapshot graphs
with \num{17} iterations each, which results in
update sequences between \num{1.6}\Mega{} and \num{702}\Mega{}.
As they are constant in size, there are roughly equally many insertions and
deletions.
Their densities vary between \num{0.7} and \num{16.4}.
The instances in \texttt{kronecker-growing} were created from thirteen snapshot
graphs with five up to \num{17} iterations, resulting in \num{282}\Kilo{} to
\num{82}\Mega{} update operations, \SI{66}{\percent} to \SI{75}{\percent} of
which are insertions.
Their densities are between \num{0.9} and \num{16.4}.

\emph{Real-World Instances.}
Our set of instances comprises all six directed, dynamic instances available
from the Koblenz Network Collection \KONECT{}~\cite{konect}, which correspond
to the hyperlink network of Wikipedia articles for six different languages.
In case of dynamic graphs, the update sequence is part of the instance.
However, the performance of algorithms may be affected greatly if an originally
real-world update sequence is permuted randomly~\cite{HHS20}.
For this reason, we also consider five ``shuffled'' versions per real-world
network, where the edge insertions and deletions have been permuted randomly.
We refer to the set of original instances as \texttt{konect} and to the
modified ones as \texttt{konect-shuffled}.
\Table{realworld-instances} lists the detailed numbers for all real-world
instances and the respective average values for the shuffled instances.
In each case, the updates are dominated by insertions.
\subsection*{Experimental Results}
We ran the algorithms \SV{}, \SVA{}, and \SVC{} with different parameters:
For \SVp{$k$}, we looked at $k=1$, $k=2$, and $k=3$, which pick one,
two, and three supportive vertices during initialization, respectively, and
never reconsider this choice.
We evaluate the variant that periodically picks new supportive vertices,
\SVAp{$k, c$}, with $k=1$ and  $c = 1\Kilo{}$, $c = 10\Kilo{}$, and $c =
100\Kilo{}$.
Preliminary tests for \SVC{} revealed $z = \num{25}$ as a good threshold for
the minimum \SCC{} size on smaller instances and $z=\num{50}$ on larger.
The values considered for $c$ were again %
$10\Kilo{}$ and $100\Kilo{}$.
\FBSBFS{} served as fallback for all \AlgLongName{Supportive Vertices}
algorithms.
Except for random ER instances with $n=\num{10}\Mega{}$, all experiments also
included \SBFS{}, \SDFS{}, and \SDBFS{}; to save space, the bar plots only show
the result for the best of these three.
We used \SES{} as subalgorithm on random instances, and \SIM{} on real-world
instances, in accordance with the experimental study for single-source
reachability~\cite{HHS20}.
\input{figures_plot-random100k-111-reduced}
\subparagraph*{ER Instances.}
We start by assessing the average performance of all algorithms by
looking at their running times on random ER instances.
For \emph{$n = 100\Kilo{}$} and equally many insertions, deletions, and queries,
\Figure{random-100k-111-red} shows the mean running time needed to process all
updates, all queries, and their sum, all operations, absolutely as well as
the relative performances for all operations, where the mean is
taken over \num{20} instances per density.
Additional relative plots are given in \Figure{random-100k-111}.
As there are equally many insertions and deletions, the density remains
constant.
Note that all plots for random ER instances use logarithmic axes in both
dimensions.

It comes as no surprise that \SVp{$2$} and \SVp{$3$} are two and three times
slower on \emph{updates}, respectively, than
\SVp{$1$}~(cf.~\textbf{\Figure{random-100k-111-red-up-abs}}).
As their update time consists solely of the update times of their \SSoSiR{}
data structures, they inherit their behavior and become faster,
the denser the instance~\cite{HHS20}.
The additional work performed by \SVAp{$1, 1\Kilo{}$} and \SVAp{$1,
10\Kilo{}$}, which re-initialized their \SSoSiR{} data
structures $66$ and six times, respectively, is plainly visible and increases
also relatively with growing number of edges, which fits the theoretical
\mbox{(re-)}initialization time of $\bigO(n+m)$. %
Computing the \SCC{}s only initially, as \SVCp{$25,\infty$} does, led to higher
update times on very sparse instances due to an increased number of supportive
vertices, but matched the performance of \SVp{$1$} for $d \geq \num{2.5}$.
As expected, re-running the
\SCC{} computation negatively affects the update time.
In contrast to \SVAp{$1,10\Kilo{}$}, however, \SVCp{$25,10\Kilo{}$} keeps a
supportive vertex as long as it still represents an \SCC{}, and thereby saves
the time to destroy the old \SSoSiR{} data structures and
re-initialize the new ones.
Evidently, both \SVA{} algorithms used a single supportive vertex for $d \geq
\num{2.5}$.

Looking at \emph{queries}
(cf.~\textbf{\Figure{random-100k-111-red-query-abs}}),
it becomes apparent that \SVCp{$25,10\Kilo{}$} can make use of its well-updated
\SCC{} representatives as supportive vertices and speed up queries up to a
factor of \num{54} in comparison to \SVA{} and \SV{}.
Up to $d=\num{3}$, it also outperforms \SVCp{$25,\infty$}.
For larger densities, the query times among all dynamic algorithms level up
progressively and reach at least equality already at $d=\num{2.5}$ in case of
\SVp{$2$} and \SVp{$3$}, at $d=\num{5}$ in case of \SVp{$1$}, and at
$d=\num{10}$ at the latest for all others. %
This matches a well-known result from random graph theory that simple ER graphs
with $m > n \ln n$ are strongly connected with high
probability~\cite{Bollobas01}.
The running times also fit our investigations into the mean percentage of queries
answered during the different stages in \Routine{query($\cdot$)} by the
algorithms (see also \Figure{random-100k-111-stats}):
For $d=\num{2}$, \SVp{$1$} could answer \SI{80}{\percent} of all queries
without falling back to \FBSBFS{}, which grew to almost \SI{100}{\percent}
for $d=\num{5}$ and above.
\SVp{$2$} answered even more than \SI{95}{\percent} queries without fallback
for $d=\num{2}$, and close to \SI{100}{\percent} already for $d=\num{3}$.
The same applied to \SVCp{$25,\infty$}, which could use \SCC{}
representatives in the majority of these fast queries.
\SVp{$1$} and \SVp{$2$} instead used mainly observation
\LabelEnumi{(O\ref{item:in-s-out-t})}, but also
\LabelEnumi{(O\ref{item:out-s-nout-t})} and
\LabelEnumi{(O\ref{item:nin-s-in-t})} in up to \SI{10}{\percent} of all queries.
As all vertices are somewhat alike in ER graphs, periodically picking new
supportive vertices does not affect the mean query performance.
In fact, \SV{} and \SVA{} are up to \SI{20}{\percent} faster than \SVC{} on the
medium and denser instances, which can be explained by the missing overhead for
maintaining the map of representatives.
All \AlgLongName{Supportive Vertices} algorithms process queries considerably
faster than \FBSBFS{}.
The average speedup ranges between almost \num{7} on the sparsest graphs in
relation to \SVCp{$25,10\Kilo{}$} and more than \num{240} in relation to
\SVp{$1$} on the densest ones.
The traditional static algorithms \SBFS{}, \SDFS{}, as well as the hybrid
\SDBFS{} were \emph{distinctly slower} by a factor of up to \num{31}\Kilo{}
(\SBFS{}) and almost \num{70}\Kilo{} (\SDFS{}, \SDBFS{}) in comparison to
\SVp{$1$}, and even \num{53} to \num{130} and \num{290} times slower than \FBSBFS{}.

In sum over \emph{all operations}, if there are equally many insertions,
deletions, and queries
(cf.~\textbf{\FiguresSub{random-100k-111-red-ops-abs}{random-100k-111-red-ops-rel}}),
\SVCp{$25,\infty{}$} and \SVp{$1$} were the \emph{fastest algorithms} on all
instances, where \SVCp{$25,\infty{}$} won on the sparser and \SVp{$1$} won on
the denser instances.
For $d=\num{1.25}$, \FBSBFS{} was almost as fast, but up to \num{45} times
slower on all denser instances.
\SVp{$2$} and \SVp{$3$} could not compensate their doubled and tripled
update costs, respectively, by their speedup in query time,
which also holds for \SVCp{$25,10\Kilo{}$}.
\SBFS{}, \SDFS{}, and \SDBFS{} were between \num{54} and \num{13}\Kilo{}
times slower than \SVCp{$25,\infty{}$} and \SVp{$1$}, despite the high
proportion of updates, and are therefore \emph{far from competitive}.

\input{figures_plot-random100k-112_random10m-111}
We repeated our experiments with $n = 100\Kilo{}$ and \emph{\SI{50}{\percent}
queries} among the operations and equally many insertions and deletions, as
well as with \emph{$n = 10\Mega{}$} and equal ratios of insertions, deletions, and
queries.
The results, shown in \textbf{\Figure{random-100k-112_10m}}, \emph{confirm} our
findings above.
In case of \emph{\SI{50}{\percent} queries}, a second supportive vertex as in
\SVp{$2$} additionally stabilized the mean running time in comparison to
\SVp{$1$}, up to
$d=\num{5}$~(cf.~\textbf{\FiguresSub{random-100k-112-ops-abs}{random-100k-112-ops-rel}}),
but none of them could beat \SVCp{$25,\infty$} on sparse instances.
On denser graphs, \SVp{$1$} was again equally fast or even up to
\SI{20}{\percent} faster.
As expected due to the higher ratio of queries, \FBSBFS{} lost in
competitiveness in comparison to the above results and is between \num{1.6} and
almost \num{80} times slower than \SVCp{$25,\infty$} on dense instances.
On the set of larger instances with \emph{$n =
10\Mega{}$}~(cf.~\textbf{\FiguresSub{random-10m-111-ops-abs}{random-10m-111-ops-rel}}),
\SVAp{1,1\Kilo{}} reached the timeout set at \SI{2}{h} on instances with $d
\geq \num{20}$.
The \emph{fastest} algorithms on average across all densities were again
\SVp{$1$} and \SVCp{$50,\infty$}.
\FBSBFS{} won for $d =\num{1.25}$, where it was about \SI{20}{\percent} faster
than \SVCp{$50,\infty$} and \SI{10}{\percent} faster than \SVp{$1$}.
Its relative performance then deteriorated with increasing density up to a
slowdown factor of \num{91}.
Except for $d =\num{1.25}$, \SVCp{$50,\infty$} outperformed \SVp{$1$} on very
sparse instances and was in general also more stable in performance, as can be
observed for $d=\num{8}$:
Here, \SVp{$1$} picked a bad supportive vertex on one of the instances, which
resulted in distinctly increased mean, median, and maximum query times.
On instances with density around $\ln n$ and above, \SVp{$1$} was slightly
faster due to its simpler procedure to answer queries and also more stable
than on sparser graphs.

In \emph{summary}, \SVC{} with $c=\infty$ clearly showed the \emph{best and
most reliable performance} on average, closely followed by \SVp{$1$}, which was
\emph{slightly faster} if the graphs were \emph{dense}.
\begin{figure}[tb]
\centering
\begin{tikzpicture}[inner sep=0pt]
\node[inner sep=0pt] (kronecker-short) {\includegraphics[width=\linewidth]{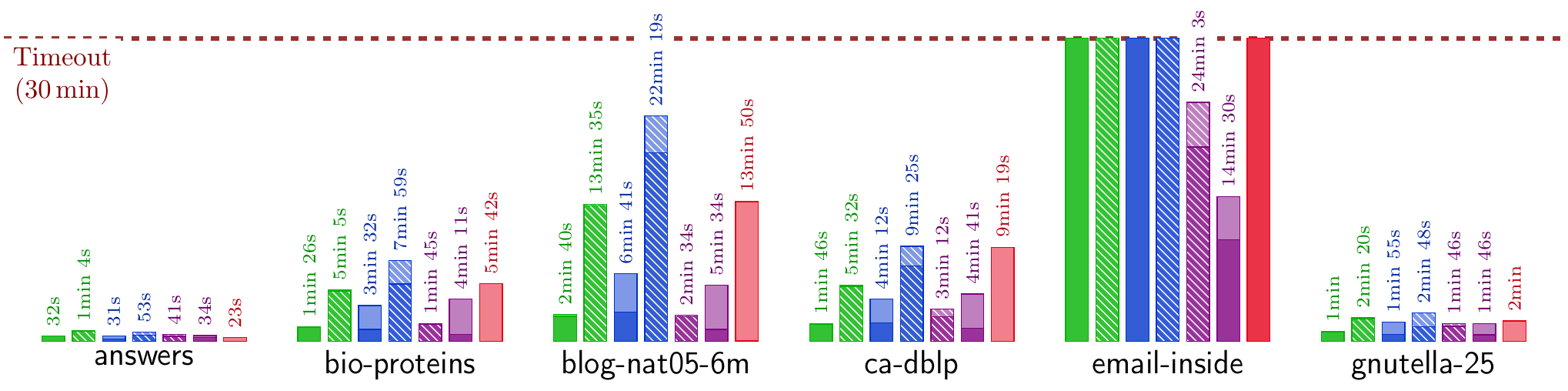}};
\node[inner sep=0pt,anchor=south] (kronecker-short-legend) at (kronecker-short.north) {\includegraphics[width=\linewidth]{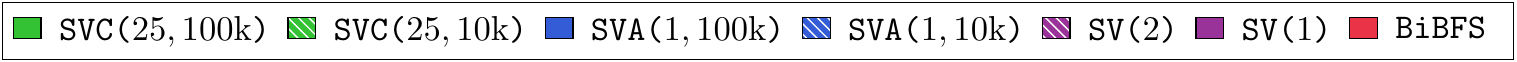}};
\end{tikzpicture}
\caption{Total update (dark color) and query (light color) times on selected \texttt{kronecker-csize} instances.
All results are shown in \Figures{kronecker-csize}{kronecker-csize-fast}.}%
\label{fig:kronecker-csize-short}
\end{figure}

\subparagraph*{Kronecker Instances.}
In contrast to ER instances, stochastic Kronecker graphs were designed to model
real-world networks, where vertex degrees typically follow a power-law
distribution and the neighborhoods are more diverse.
For this reason, the choice of supportive vertices might have more effect on
the algorithms' performances than on ER instances.
\textbf{\Figure{kronecker-csize-short}} shows six selected results for all dynamic
algorithms as well as \FBSBFS{}
on \texttt{kronecker-csize} instances
with a query ratio of
\SI{33}{\percent} (see \Figures{kronecker-csize}{kronecker-csize-fast} for
the complete set):
Each bar consists of two parts, the darker one depicts the total update time,
the lighter one the total query time, which is comparatively small for most
dynamic algorithms and therefore sometimes hardly discernible.
In case that an algorithm reached the timeout, we only use the darker
color for the bar.
The description next to each bar states the total operation time.
By and large, the picture is very similar to that for ER instances.
On \num{13} and \num{14} out of the \num{20} instances, \SBFS{}/\SDBFS{} and
\SDFS{}, respectively, did not finish within six hours.
As on the previous sets of instances, these algorithms are \emph{far from
competitive}.
The performance of \FBSBFS{} was ambivalent:
it was the fastest on two instances, but lagged far behind on others.
\SVp{$1$} and \SVp{$2$} showed the \emph{best
performance} on the \emph{majority} of instances and
were the \emph{only ones} to finish within six hours on \textsf{email-inside}.
On some graphs, the total operation time of \SVp{$1$} was dominated by the
query time, \eg, on \textsf{bio-proteins}, whereas \SVp{$2$} was able to reduce
the total operation time by more than half by picking a second supportive
vertex.
However, \SVCp{$25,100\Kilo{}$} was even able to outperform this slightly and
was the \emph{fastest} algorithm on half of the instances.
As above, recomputing the \SCC{}s more often (\SVCp{$25,10\Kilo{}$}) or
periodically picking new support vertices (\SVAp{$1,1\Kilo{}$},
\SVAp{$1,10\Kilo{}$}) led to a slowdown in general.

On \texttt{kronecker-growing}, \SVp{$1$} was the \emph{fastest} algorithm on
all but one instance.
The overall picture is very similar, see
\Figures{kronecker-growing}{kronecker-growing-fast}.
\input{figures_plot-konect-original}
\subparagraph*{Real-World Instances.}
In the same style as above, \textbf{\Figure{konect-original}} shows the results
on the six real-world instances with real-world update sequences,
\texttt{konect}, again with \SI{33}{\percent} queries among the operations.
We set the timeout at \SI{24}{h}, which was reached by \SBFS{}, \SDFS{}, and
\SDBFS{} on all but the smallest instance.
On the largest instance, \texttt{DE}, they were able to process only around
\SI{6}{\percent} of all updates and queries within this time. %
The \emph{fastest algorithms} were \SVp{$1$} and \SVp{$2$}.
If \SVp{$1$} chose the single support vertex well, as in case of \texttt{FR},
\texttt{IT}, and \texttt{SIM}, the query costs and the total operation times
were low; on the other instances, the second support vertex, as chosen by
\SVp{$2$}, could speed up the queries further and even compensate the cost
for maintaining a second pair of \SSoSiR{} data structures.
Even though the instances are growing and most vertices were isolated and
therefore not eligible as supportive vertex during initialization,
periodically picking new supportive vertices, as \SVA{} does, did not
improve the running time.
\SVCp{$50,\infty$} performed well, but the extra effort to \emph{compute the
\SCC{}s} and use their representatives as supportive vertices \emph{did not pay
off}; only on \texttt{SIM}, \SVCp{$50,\infty$} was able to outperform both
\SVp{$1$} and \SVp{$2$} marginally.

Randomly permuting the sequence of update operations, as for the instance set
\texttt{konect-shuffled}, did not change the overall picture.
The results are shown in \Figure{konect-shuffled}.
\section{Conclusion}
Our extensive experiments on a diverse set of instances draw a somewhat
surprisingly consistent picture:
The most simple algorithm from our family, \SVp{$1$}, which picks a single
supportive vertex, performed extremely well and was the fastest on a large
portion of the instances.
On those graphs where it was not the best, \SVp{$2$} could speed up the running
time by picking a second supportive vertex.
Additional statistical evaluations showed that already for sparse graphs,
\SVp{$1$} and \SVp{$2$} answered a great majority of all queries in constant
time using only its supportive vertices.
Recomputing the strongly connected components of the graph
in very large intervals and using
them for the choice of supportive vertices yielded a comparatively good or
marginally better algorithm on random instances, but not on real-world graphs.

The classic static algorithms BFS and DFS, which were competitive or even
superior to the dynamic algorithms evaluated experimentally in previous studies,
lagged far behind the new algorithms and were outperformed by several orders of
magnitude.
\bibliography{paper}
\clearpage

\renewcommand{\thefigure}{\thesection.\arabic{figure}}
\renewcommand{\thetable}{\thesection.\arabic{table}}
\begin{appendix}
\section{Appendix}
\subsection{Related Work}%
\label{sect:app:related}
Over the years, extensive research on the (fully) dynamic transitive closure
problem has spawned a large body of algorithms.
They can be classified with respect to at least the following five criteria:
the dynamic setting (incremental, decremental, or fully),
the supported input instances (arbitrary or \DAG{}s),
their mode of operation (deterministic, Las Vegas-style randomized, or Monte
Carlo-style randomized),
whether they are combinatorial or rely on fast matrix multiplication,
and whether they support queries in constant time or not.
Note that Monte Carlo algorithms may answer with one-sided error with some
(small) probability, whereas Las Vegas algorithms always answer correctly, but
their running time may depend on random variables.

The insertions-only case was first considered in 1983 by Ibaraki and
Katoh~\cite{IK83}, whose algorithm processes any number of insertions in
$\bigO(n^3)$ total time with $\bigO(1)$ query time.
Independently from each other, Italiano~\cite{Italiano86} as well as
La Poutré and van Leeuwen~\cite{LL88} improved this result to $\bigO(n)$
amortized insertion time.
Yellin's algorithm~\cite{Yellin93} can process a sequence of edge insertions on
an initially empty graph in $\bigO(\tcE{} \cdot \maxOutDegree{})$ total time,
where $\tcE{}$ is the number of edges in the transitive closure and
$\maxOutDegree{}$ the maximum outdegree of the final graph.

In the deletions-only case, Ibaraki and Katoh~\cite{IK83} gave the first
algorithm in 1983 with constant query time and $\bigO(n^2(m+n))$ time for any
number of deletions, which was again improved by La Poutré and van
Leeuwen~\cite{LL88} to $\bigO(m)$ amortized deletion time.
An algorithm by Frigioni~\ea{}~\cite{FMNZ01} achieves the same and uses
$\bigO(n\cdot m)$ time for initialization.
Demetrescu and Italiano's non-combinatorial algorithm~\cite{DI08} reduced the
amortized deletion time further to $\bigO(n)$, but requires $\bigO(n^3)$
preprocessing time.
Earlier, a Monte Carlo algorithm by Henzinger and King~\cite{HK95} was able to
reduce the amortized deletion time to $\bigO(n\log^2 n)$, however at the
expense of $\bigO(\frac{n}{\sqrt{n}})$ query time.
Baswana~\ea{}~\cite{BHS02} showed with another Monte Carlo algorithm
that the query time can be kept constant with
$\bigO(n^{\frac{4}{3}}\sqrt[3]{\log n})$ amortized deletion time.
A Las Vegas algorithm by Roditty and Zwick~\cite{RZ08} improved the total
running time to $\bigO(n\cdot m)$, and \L{}\c{a}cki~\cite{Lacki11} finally showed
in 2011 that the same can be achieved with a deterministic algorithm.
On \DAG{}s, Italiano~\cite{Italiano88} presented the first algorithm with
constant query time and an amortized deletion time of $\bigO(n)$ already in
1988, and Yellin~\cite{Yellin93} gave an algorithm with a total deletion time
of $\bigO(\tcE{} \cdot \maxOutDegree{})$, where $\tcE{}$ and $\maxOutDegree{}$
are defined as above.

Henzinger and King~\cite{HK95} gave the first fully dynamic algorithms for
transitive closure, which are Monte Carlo-style randomized with
$\bigO(\frac{n}{\sqrt{n}})$ query time and either $\bigO(m\sqrt{n}\log^2 n)$
or $\bigO(m^{0.58}\cdot n)$ amortized update time.
King and Sagert~\cite{KS02} improved upon these results using a Monte Carlo
algorithm with $\bigO(n^{2.26})$ amortized update time, constant query time,
and $\bigO(n^2)$ space.
Whereas the aforementioned results rely on fast matrix multiplication, a
combinatorial algorithm by King~\cite{King99} reduced the amortized update time
further to $\bigO(n^2\log n)$.
King and Thorup~\cite{KT01} decreased the space complexity of King's
algorithm~\cite{King99} from $\bigO(n^3)$ to $\bigO(n^2\log n)$.
Using non-combinatorial techniques, Demetrescu and Italiano~\cite{DI08} showed
that the amortized update time can be reduced further to $\bigO(n^2)$.
In the following, Roditty and Zwick~\cite{RZ08} presented a deterministic,
combinatorial algorithm with $\bigO(\sqrt{n})$ query time and
$\bigO(m\sqrt{n})$ amortized update time as well as a non-combinatorial Monte
Carlo algorithm with $\bigO(m^{0.43})$ query time and $\bigO(m^{0.58}n)$
amortized update time.
Sankowski~\cite{Sankowski04} gave three non-combinatorial Monte Carlo
algorithms with constant, $\bigO(n^{0.58})$, or $\bigO(n^{1.495})$ worst-case
query time and $\bigO(n^2)$, $\bigO(n^{1.58})$, or $\bigO(n^{1.495})$
worst-case update time, respectively.
For constant query time, Roditty~\cite{Roditty08} developed a deterministic,
combinatorial algorithm with $\bigO(n^2)$ amortized update time,
which in addition reduces the initialization time from $\bigO(n^3)$~\cite{DI08}
and above to $\bigO(n\cdot m)$.
Roditty and Zwick~\cite{RZ16} showed with a deterministic, combinatorial
algorithm that the amortized update time can be reduced down to $\bigO(m+n\log
n)$ at the expense of $\bigO(n)$ worst-case query time.
Recently, van den Brand~\ea{}~\cite{BNS19} presented an algorithm with
$\bigO(n^{1.407})$ worst-case update and query time.

For \DAG{}s, King and Sagert~\cite{KS02} gave a Monte Carlo algorithm with
constant query time and $\bigO(n^2)$ amortized update time.
Another Monte Carlo algorithm by Demetrescu and Italiano~\cite{DI05} reduced
the update time to $\bigO(n^{1.575})$ even in worst case, however at the
expense of a $\bigO(n^{0.575})$ worst-case query time.
Roditty and Zwick~\cite{RZ08} presented a deterministic algorithm with
$\bigO(\frac{n}{\log n})$ query time and $\bigO(m)$ amortized update time.
Roditty~\cite{Roditty08} showed that the transitive closure can be maintained
in the fully dynamic setting on \DAG{}s using $\bigO(n^2)$ space and with
$\bigO(n\cdot m)$ initialization time, $\bigO(n^2)$ amortized insertion time,
constant amortized deletion time, and constant query time.

These results are counteracted by the fact that the insertion or deletion of a
single edge may affect the reachability between $\Omega(n^2)$ pairs of
vertices.
If queries shall be answered in constant time, the pairwise reachabilities have
to be stored explicitly, which in turn implies that a worst-case update time of
$\bigO(n^2)$ is the best one can hope for.
So far, this time bound has only been reached in the amortized cost
model~\cite{Roditty08}.
Also note that an incremental algorithm with $\bigO(n^{1-\varepsilon})$ query
time and $\bigO(n^{1-\varepsilon})$ insertion time per edge or
$\bigO(n^{2-\varepsilon})$ insertion time per vertex along with its incident
edges would immediately yield an improved static algorithm for transitive
closure.
Moreover, Henzinger~\ea{}~\cite{HKNS15} have shown that unless the Online
Matrix-Vector Multiplication problem can be solved in
$\bigO(n^{3-\varepsilon})$, $\varepsilon > 0$, no partially dynamic algorithm for
transitive closure exists with polynomial preprocessing time,
$\bigO((mn)^{1-\delta})$ update time, and $\bigO(m^{\delta'-\delta})$ query time
simultaneously, for $\delta' \in (0,\frac{1}{2}]$, $m \in
\Theta(n^{\frac{1}{1-\delta'}})$, $\delta > 0$.
On condition of a variant of the OMv conjecture, van den
Brand~\ea{}~\cite{BNS19} improved the lower bounds for update and query time to
$\Omega(n^{1.406})$.
Slightly weaker lower bounds, which are additionally restricted to
combinatorial algorithms, can be achieved based on the Boolean Matrix
Multiplication conjecture~\cite{DHZ00}.

In large studies, Frigioni~\ea{}~\cite{FMNZ01} as well as Krommidas and
Zaroliagis~\cite{KZ08} have implemented an extensive set of the above
mentioned algorithms and compared them to each other as well as
to static, ``simple-minded'' algorithms such as breadth-first and depth-first
search.
The set comprises the algorithms by
Italiano~\cite{Italiano86,Italiano88}, Yellin~\cite{Yellin93}, Henzinger and
King~\cite{HK95}, King~\cite{King99}, King and Thorup~\cite{KT01},
Roditty~\cite{Roditty08}, Roditty and Zwick~\cite{RZ08,RZ16}, Demetrescu and
Italiano~\cite{DI05,DI08}, as well as several modifications and improvements
thereof.
The experimental evaluations on random Erd\H{o}s-Renyí graphs, instances
constructed to be difficult on purpose, as well as two instances based on
real-world graphs, showed that on all instances except for dense random graphs
or a query ratio of more than 65\%
dynamic ones distinctly and up to several factors.
Their strongest competitors were the fully dynamic extension~\cite{FMNZ01} of
the algorithms by Italiano~\cite{Italiano86,Italiano88}, as well as the fully
dynamic extension~\cite{KZ08} of the decremental algorithm by Roditty and
Zwick~\cite{RZ08}.
These two algorithms also were the only ones that were faster than
static graph traversal on dense random graphs.

\clearpage
\subsection{Additional Tables and Plots}
\begin{figure}[htb]
\centering

\includegraphics[width=4.5cm]{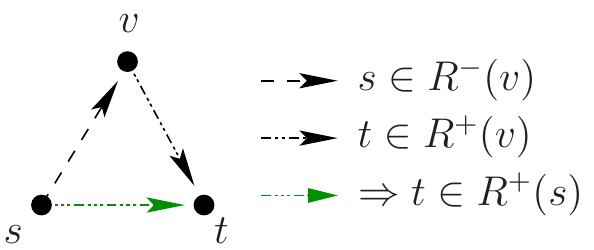}
\includegraphics[width=4.5cm]{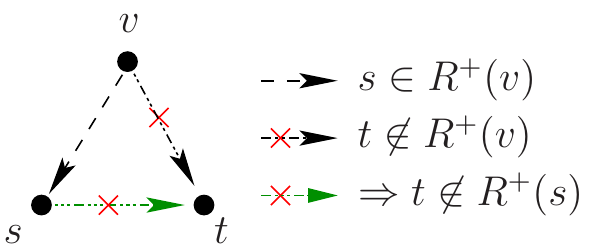}
\includegraphics[width=4.5cm]{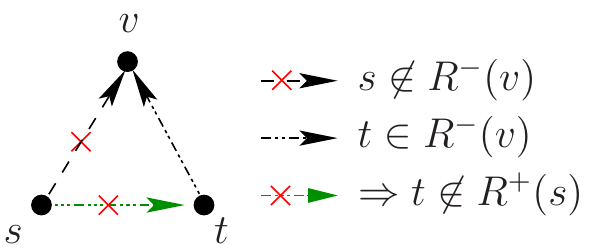}
\caption{Basic observations \LabelEnumi{(O\ref{item:in-s-out-t})},
                            \LabelEnumi{(O\ref{item:out-s-nout-t})},
                            \LabelEnumi{(O\ref{item:nin-s-in-t})}
														that help to decide reachability quickly.}
\label{fig:observations}
\end{figure}

\begin{table*}[htb]
\caption{Worst-case running times and space requirements.}
\TabLabel{algorithms}
\centering
\astr{1.2}
\begin{tabular}{@{}l@{\hspace{2em}}ccc@{\hspace{2em}}cc@{}}
\toprule
&
&
Time
&
&
\multicolumn{2}{c}{Space}
\\
Algorithm & Insertion & Deletion & Query & Permanent & Update\\
\midrule
\makecell[cl]{%
\SBFS{},
\SDFS{},\\
\SDBFS{},
\FBSBFS{}
}
&
$0$
&
$0$
&
$\bigO(n+m)$
&
$0$
&
$\bigO(n)$
\\
\makecell[cl]{%
\SVp{$k$}/\SVAp{$k,c$}\\
$\llcorner$ \qquad with \SIM{}\\
$\llcorner$ \qquad with \SES{}\\
}
&
\makecell[cc]{%
~\\
$\bigO(k \cdot (n + m))$
}
&
\makecell[cc]{%
~\\
$\bigO(k \cdot n \cdot m)$\\
$\bigO(k \cdot (n + m))$\\
}
&
\makecell[cc]{%
~\\
$\bigO(k + (n+m))$
}
&
\makecell[cc]{%
~\\
$\bigO(k \cdot n)$
}
&
\makecell[cc]{%
~\\
$\bigO(n)$
}
\\
\makecell[cl]{%
\SVCp{$z,c$}\\
$\llcorner$ \qquad with \SIM{}\\
$\llcorner$ \qquad with \SES{}\\
}
&
\makecell[cc]{%
~\\
$\bigO(n \cdot (n + m))$
}
&
\makecell[cc]{%
~\\
$\bigO(n^2 \cdot m)$\\
$\bigO(n \cdot (n + m))$\\
}
&
\makecell[cc]{%
~\\
$\bigO(n+m)$
}
&
\makecell[cc]{%
~\\
$\bigO(n^2)$
}
&
\makecell[cc]{%
~\\
$\bigO(n)$
}
\\
\bottomrule
\end{tabular}
\end{table*}

\begin{table*}[h]
\caption{%
Number of vertices $n$,
initial, average, and final number of
edges
$\initE$, $\avgE$, and $\finE$,
average density $\avgD$,
total number of updates $\Updates$,
and percentage of additions $\AddsRatio$
among updates, %
of real-world instances.}
\TabLabel{realworld-instances}
\centering
\astr{1.2}
\begin{tabular}{@{}lrrrrrrrrr@{}}
\toprule
Instance & $n$ & $\initE$ & $\avgE$ & $\finE$ & $\avgD$ & $\Updates$ & $\AddsRatio$ \\
\midrule
\texttt{FR}           & \num{2.2}\Mega &     \num{3} & \num{13.0}\Mega & \num{24.5}\Mega &  \num{5.9} & \num{59.0}\Mega{} & \SI{71}{\percent} \\
\texttt{DE}           & \num{2.2}\Mega &     \num{4} & \num{16.7}\Mega & \num{31.3}\Mega &  \num{7.7} & \num{86.2}\Mega & \SI{68}{\percent}  \\
\texttt{IT}           & \num{1.2}\Mega &     \num{1} &  \num{9.3}\Mega & \num{17.1}\Mega &  \num{7.8} & \num{34.8}\Mega & \SI{75}{\percent}  \\
\texttt{NL}           & \num{1.0}\Mega &     \num{1} &  \num{5.7}\Mega & \num{10.6}\Mega &  \num{5.4} & \num{20.1}\Mega & \SI{76}{\percent}  \\
\texttt{PL}           & \num{1.0}\Mega &     \num{1} &  \num{6.6}\Mega & \num{12.6}\Mega &  \num{6.4} & \num{25.0}\Mega & \SI{75}{\percent}  \\
\texttt{SIM}          &  \num{100}\Kilo &     \num{2} &   \num{401}\Kilo &   \num{747}\Kilo &  \num{5.9} &  \num{1.6}\Mega & \SI{73}{\percent}  \\
\midrule
\texttt{FR\_SHUF}     & \num{2.2}\Mega &   \num{4.0} & \num{16.4}\Mega & \num{30.4}\Mega & \num{7.4} & \num{53.1}\Mega & \SI{79}{\percent} \\
\texttt{DE\_SHUF}     & \num{2.2}\Mega &   \num{3.8} & \num{22.6}\Mega & \num{41.1}\Mega & \num{10.4} & \num{76.4}\Mega & \SI{77}{\percent} \\
\texttt{IT\_SHUF}     & \num{1.2}\Mega &   \num{3.8} & \num{10.9}\Mega & \num{20.5}\Mega & \num{9.1} & \num{31.4}\Mega & \SI{83}{\percent} \\
\texttt{NL\_SHUF}     & \num{1.0}\Mega &   \num{3.8} & \num{6.7}\Mega  & \num{12.6}\Mega  & \num{6.4} & \num{18.1}\Mega & \SI{85}{\percent} \\
\texttt{PL\_SHUF}     & \num{1.0}\Mega  &  \num{3.6} & \num{7.9}\Mega & \num{14.9}\Mega & \num{7.7} & \num{22.7}\Mega & \SI{83}{\percent} \\
\texttt{SIM\_SHUF}    & \num{100}\Kilo & \num{5.6} & \num{476}\Kilo & \num{892}\Kilo & \num{4.7} & \num{1.6}\Mega & \SI{80}{\percent} \\
\bottomrule
\end{tabular}
\end{table*}

\input{figures_plot-random100k-111}
\input{figures_plot-random100k-111-stats}
\input{figures_plot-kronecker-csize}
\input{figures_plot-kronecker-csize-fast}
\input{figures_plot-kronecker-growing}
\input{figures_plot-kronecker-growing-fast}
\input{figures_plot-konect-shuffled}
\end{appendix}

\end{document}